\documentstyle[12pt]{article}
\advance\hoffset by -1.1truecm
\begin{document}
\title{Simple considerations on the behaviour of bosonic modes
with quantum group symmetry}
\author{Achim Kempf\thanks{supported by Studienstiftung des
deutschen Volkes, BASF-Fellow}
\\ Department of Applied Mathematics \& Theoretical Physics\\
University of Cambridge\\ Cambridge CB3 9EW, U.K.}
\date{ }
\maketitle
\vskip-8.5cm
\hskip4.5cm
\tt preprint DAMTP/93-43 \quad and \quad hep-th/9308025  \rm
\vskip8cm
\begin{abstract}
While it is possible to introduce quantum group symmetry into
the framework of quantum mechanics, the general problem
of how to implement quantum group symmetry into $(3+1)$ dimensional
quantum field theory has not yet been solved. Here we try to estimate
some features of the behaviour of bosonic modes.

\end{abstract}

\section{Introduction}
Quantum groups arose from the quantum inverse scattering method
 \cite{frt}.
Mathematically they are deformations of groups \cite{DRI,maj}.
The underlying structure is that of noncommutative geometry:
The (co-quasitriangular Hopf-) algebra
 of functions on the group becomes
noncommutative when a new parameter, which is usually
denoted '$q$', deviates from $1$, see e.g. \cite{DRI,maj,SWE,ABE}.

Quantum groups have found a wide field of applications from
statistical mechanics to knot theory
, see e.g. \cite{devega,kauffman}
and are also extensively
studied in their own right, see e.g. \cite{mleute,SUD,MAN2,wz,akny}.
Particularly tempting is the idea to
implement quantum groups as symmetries in quantum theory, which
 essentially
means to make the commutation relations $q$- dependent.
The usual results are then recovered as the special case $q=1$.

The case $q\ne 1$ may e.g. apply to the description of
non-standard commutation relations of collective excitations or
quasi-particles in solid state physics. But one can also consider
 $q \ne 1$
as a check for the usual quantum theory. In this context,
the observation of discretizing and regularizing effects of quantum group
symmetry is of great interest \cite{cheftoy,POI1,POI2,arthur}.

\section{Commutation relations with $SU_q(n)$- symmetry}
There are already several approaches to the definition of
quantum mechanical commutation relations with quantum group symmetry
in the literature,
see e.g. \cite{woro,bieden,macf,meinslmp,meinsjmp,kulishluk,zachos}.
In particular Heisenberg algebras
generated by adjoint pairs of operators
${a^{\dagger}}_i$ and $a^i$ were developed.
The following $q$-dependent bosonic commutation relations
are conserved under the action of the quantum group $SU_q(n)$,
see e.g. \cite{woro,meinslmp}:
\begin{eqnarray*}
a_i a_j - 1/q a_j a_i & =  & 0 \mbox{ \quad for \quad } i > j \\
a^{\dagger}_i a^{\dagger}_j - 1/q a^{\dagger}_j a^{\dagger}_i & =  & 0
\mbox{ \quad for \quad } i < j \\
a_i a^{\dagger}_j  - q \mbox{ }a^{\dagger}_j a_i & =  & 0
\mbox{ \quad for \quad } i \ne j \\
a_i a_i^{\dagger}
- q^2 a^{\dagger}_i a_i  & =  & 1
+ (q^2 - 1) \sum\limits_{j<i} a^{\dagger}_j a_j \\
\end{eqnarray*}
The ground state can be defined as usual:
$$ < 0 \vert 0 > = 1 \mbox{\quad and\quad } {a_i \vert 0 >} = 0
\mbox{\quad for\quad } i = 1,...,n $$
One then obtains for the scalar product $< , >$:
\begin{equation}
<0\vert (a_n)^{r_n} \cdot ... \cdot (a_1)^{r_1}
 (a_1^{\dagger})^{r_1} \cdot ... \cdot (a_n^{\dagger})^{r_n} \vert 0>
 \quad  =
 \prod\limits_{i=1}^n [r_i]_{q}!
\end{equation}
with
\begin{equation}
[r]_q! := [1]_q \cdot [2]_q \cdot [3]_q \cdot . . . \cdot [r]_q
 \mbox{ \qquad and \qquad }
[p]_q := \frac{q^{2p} - 1}{q^{2} - 1}
\label{norm}
\end{equation}
Although quantum groups do in general have more than one free parameter,
no further parameters enter in these commutation relations \cite{berlin}.
The usual
quantum mechanical programme, representation on a positive definit
(Bargmann Fock-) Hilbert space of
 wave functions and definition of integral
 kernels like Green functions etc., could be performed \cite{meinsjmp}, while
 preserving the hermiticity of the
observables and the unitarity of time evolution.
\section{q-bosonic modes}
An extension of this formalism to relativistic
quantum field theories proves to be
difficult \cite{meinsqft}. However, already by using
simple nonrelativistic quantum mechanical techniques one can
try to obtain some insight into the behaviour of modes of bosonic
fields with quantum group symmetry. The reason for this is the estimation,
that effects of $q\ne 1$ do not primarily occur as high energy effects,
but instead as effects of high occupation numbers - at least on the basis
of the quantum mechanical formalism mentionend above.
\medskip\newline
To see this, let us consider the simple quantum mechanical Hamiltonian
of the $n$- dimensional isotropic harmonic oscillator:
$$H := \hbar \omega \sum_{i=1}^{n}
a^{\dagger}_i a_i \mbox{\qquad with } \omega \in {I\!\!R}^+$$
In the $SU_q(n)$ symmetric formalism it has the energy
spectrum \cite{meinslmp}:
\begin{equation}
E_p = \hbar \omega (0 + 1 + q^2 + q^4 + ... + q^{2(p-1)}) =
\hbar \omega \frac{q^{2p} - 1}{q^{2} - 1} \mbox{ \quad with \quad }
p = 0,1,2,...
\label{spec}
\end{equation}
Thus its energy levels are no longer equidistant. The spacing of the levels
increases or decreases
depending on wether $q$ is larger or smaller than $1$. In the latter case
the series actually converges and the spectrum gets an upper bound
$\frac{\hbar \omega}{1 - q^2}$. In both cases it is true that
the higher the energy level is, the more it deviates from the corresponding
energy level for $q=1$.
\medskip
\newline
In quantum field theory there are harmonic oscillators at the points in
$3$- momentum space.
Thus, if it is possible to extend this formalism to quantum field theory,
it can be expected that quantum group effects do again increase with
the level of excitation, i.e. then with the number of field quanta that are
present.
\medskip

Let us assume in the following, that we have a
photon- or phonon- like bosonic quantum field and, for simplicity, that the
momentum space is discretized. Apart from solids or from the situation
in
 cavity,
there are also hints, that some kind
of discretization can occur naturally through quantum group symmetry
 \cite{cheftoy,POI1,POI2,arthur}.
\smallskip
\newline
Picking out a mode with fixed wave vektor $\vec{k}$, we do not need to specify
the dispersion relation in order to write its free Hamiltonian in the form
just given above:
$$H_{\vec{k}} := \hbar {\omega}_{\vec{k}} \sum_{i=1}^{n}
a^{\dagger}_i(\vec{k}) a_i(\vec{k}) $$
If we let $n=2$, the two degrees of freedom can e.g. be interpreted as two
polarizations. What follows will not depend on $n$ and
its interpretation:
\section{Induced emission of field quanta}
Let us now consider a system with a
source that is able to emit and to absorb
field quanta of the q-mode. We follow the early treatment
by Dirac \cite{dirac}.
The Hamiltonian, expanded in the $q$-mode's variables reads:
\begin{eqnarray*}
H & = & H_{source} + \sum_i (g^{\dagger}_i a_i(\vec{k}) +
a^{\dagger}_i(\vec{k}) g_{i})\\
 &   &  +  \sum_{i,j} (h_{ij} a^{\dagger}_i(\vec{k}) a_j(\vec{k})
+ f_{ij} a_i(\vec{k}) a_j(\vec{k}) + f^{\dagger}_{ij}
a^{\dagger}_i(\vec{k}) a^{\dagger}_j(\vec{k})) + ...
\end{eqnarray*}
This could e.g. be a laser-like system with an atom as the source
coupled to a photon
mode with a large wave length compared to the size of the atom. The above
Hamiltonian is then obtained from the minimal coupling of the vector
potential. In the Hamiltonian the part of the free source and the
free field
$$
H_{source} + \hbar {\omega}_{\vec{k}} a^{\dagger}_i(\vec{k})
a^{\dagger}_i(\vec{k}))
$$
are assumed to be dominant. Thus we can work in the direct product of the
Hilbert spaces of states of the free source and of the free boson mode.
Their interaction terms we then treat perturbatively.
There are terms of one-boson emission, one boson absorption,
two boson emission, etc.. The coefficients $f,g,h$ are operators
that act nontrivially only on the source states.
\smallskip

Let us now calculate for example the probability $p_m$ of the
the emission of one q-boson of the $i$'th
degree of freedom, dependent on the number $m$
of q-bosons already present. The probability amplitude ${\psi}_m$ is
proportional to the transition matrix element:
\begin{equation}
{\psi}_m \propto\quad < \mbox{source}_{after}\vert
<0\vert \frac{(a_i(\vec{k}))^{m+1}}{\sqrt{[m+1]_q!}}
\vert \quad a^{\dagger}_i g_i \quad \vert
\frac{(a^{\dagger}_i(\vec{k}))^m}{\sqrt{[m]_q!}}
\vert 0>\vert \mbox{source}_{before}>
\end{equation}
where $([m]_q!)^{-1/2}(a^{\dagger}_i(\vec{k}))^m \vert 0>$ is
a normalized state of $m$ bosons (see Eq.\ref{norm}). Further evaluation
yields:
$$
{\psi}_m \propto\quad < \mbox{source}_{after}\vert g_i
 \vert \mbox{source}_{before}>
 <0\vert \frac{(a_i(\vec{k}))^{m+1}}{\sqrt{[m+1]_q!}}
\cdot \sqrt{\frac{[m+1]_q!}{[m]_q!}} \cdot
\frac{(a^{\dagger}_i(\vec{k}))^{m+1}}{\sqrt{[m+1]_q!}} \vert 0>
$$
and finally
\begin{equation}
{\psi}_m \propto \sqrt{[m+1]_q} \mbox{ \quad  i.e. \quad }
 p_m \propto [m+1]_q.
\label{a}
\end{equation}
Thus, while the probability of spontaneous emission does not
depend on $q$, the probability of induced emission does: It is
no longer proportional to the number $m+1$, but to the q-number
$[m+1]_q$.
The deviation from the usual result increases with the
occupation number of the boson mode.

The energy contained in the boson mode is
proportional to $[m]_q$ (see Eq.\ref{spec}). Thus, while usually
for the probability of boson emission holds
\begin{equation}
p_m \propto I + const
\label{b0}
\end{equation}
with $I$ the intensity of the incident beam, we now obtain:
\begin{equation}
p_m \propto I + q^{2m} const \mbox{ \qquad i.e. \qquad }
p_m/I \propto 1 + \frac{q^{2m}(q^2-1)}{q^{2m}-1}
\label{b}
\end{equation}
The calculations for the absorption of bosons are of course analogous.
\section{Conclusion}
To summarize, from simple considerations,
based on a $SU_q(n)$-symmetric quantum mechanical
formalism, we expect that bosonic fields with quantum group symmetry
should be recognizable from their
particular behaviour for high occupation numbers $m$. Some of the
characteristics should be:

1. The probabililty of induced emission into a fixed mode is no longer
essentially proportional to its occupation number (Eq.\ref{a}).
For high occupation numbers the probability reaches a maximum if $q<1$ or,
if $q>1$, it increases exponentially.

2. On the other hand, the probability of emission is still essentially
proportional to the intensity of the beam (Eqs. \ref{b0}, \ref{b}).
For $q<1$ the intensity can only increase to a finite value.

3. Last but not least one can expect that laser-like systems of
a q-bosonic mode with a source could come 'out of tune' for
high occupation numbers. This is simply because the spacing of the energy
levels
of the mode changes with the occupation number, and the sources may only be
prepared to provide quanta of energy within a certain fixed range.


\section{References}
\begin{thebibliography}{99}
\bibitem{frt} L.D.Faddeev, N.Yu.Reshetikhin, L.A.Takhtajan,
Alg. Anal. 1,1, (1989) 178-206
\bibitem{DRI} V.G.Drinfel'd, Proceedings of the International
Congress\\ of Mathematicians, Vol. 1, 798, (1986)
\bibitem{maj} S.Majid, Int.J.Mod.Phys. A. Vol. 5, No 1 (1990) 1-91
\bibitem{SWE} M.E.Sweedler, Hopf Al\-ge\-bras, Benjamin (1969)
\bibitem{ABE} E.Abe, Hopf Al\-ge\-bras, Cambridge Univ. Press (1980)
\bibitem{devega} H.J.De Vega, Int.J.Mod.Phys. A, Vol. 4, No 10: 2371-2463
(1989)
\bibitem{kauffman} L.Kauffman, Preprint ANL-HEP-CP-90-36 (1990)
\bibitem{mleute} H.Ewen, O.Ogievetsky, J.Wess, Preprint MPI-PAE/PTh 18/91
(1991)
\bibitem{SUD} A.Sudbery, J.Phys. A. 23, L697,(1990)
\bibitem{MAN2} Yu.I.Manin, Commun. Math. Phys. 123, (1989) 163-175
\bibitem{wz} J.Wess, B. Zumino, Nucl. Phys. Proc. Suppl. 18B (1991) 302
\bibitem{akny} A.Kempf, in Proc. XX DGM- Conference, June '91, New York,
eds. S. Catto, A. Rocha (World Scientific, 1991) 546
\bibitem{cheftoy} J.Schwenk, J.Wess, Preprint MPI-PAE/PTh ss/92 (1992)
\bibitem{POI1} O.Ogievetsky, W.B.Schmittke, J.Wess, B.Zumino,
 Preprint MPI-Ph/91-98
\bibitem{POI2} M.Pillin, W.B. Schmidke, J.Wess, Preprint MPI-Ph/92-75
\bibitem{arthur} A.Hebecker, W.Weich, in Proc. XIX Symposium Ahrenshoop,
Eds. B.D{\"o}rfel, E.Wieczorek, Berlin 1992, DESY Preprint 93-013
\bibitem{woro} W.Pusz, S.Woronowicz, Rep. Math. Phys. 27 (1989) 231.
\bibitem{bieden} L.Biedenharn, J. Phys. A 22 (1989) L 873
\bibitem{macf} A.Macfarlane, J. Phys. A 22 (1989) 4581
\bibitem{meinslmp} A.Kempf, Lett. Math. Phys. 26: (1992) 1-12
\bibitem{meinsjmp} A.Kempf, J. Math. Phys., Vol. 34, No.3, (1993) 969-987
\bibitem{kulishluk} M.Chaichian, P. Kulish,
 J. Lukierski, Phys. Lett. B 262 (1991) 43
\bibitem{zachos} D.B.Fairlie, C. Zachos, preprint ANL-HEP-CP-91-28 (1991)
\bibitem{berlin} A.Kempf, in Proc. XIX Symposium Ahrenshoop in Wendisch-Rietz,
Eds. B.D{\"o}rfel, E.Wieczorek, DESY Preprint 93-013, (1992)
\bibitem{meinsqft} A.Kempf, Preprint LMU-TPW 92-26 (1992)
\bibitem{dirac} P.A.M.Dirac, The Principles of
 Quantum Mechanics, Oxford Univ. Press, 4th ed. (1958)
\end{thebibliography}
\end{document}